\documentclass[runningheads]{svmult}

\usepackage{makeidx}   
\usepackage{graphicx}  
\usepackage{multicol}  
\usepackage{physprbb}  
\makeindex             

\usepackage{amsmath}   %
\usepackage{amssymb}   
\newcommand\ks{{\rm\thinspace ks}}
\newcommand\kev{{\rm\thinspace keV}}
\newcommand\pcmsq{\hbox{$\rm\thinspace cm^{-2}$}}
\newcommand\ergps{\hbox{$\rm\thinspace erg~s^{-1}$}}
\newcommand\kevpcmsqpspsrpkev{\hbox{$\rm\thinspace keV~cm^{-2}~s^{-1}~sr^{-1}~keV^{-1}$}}
\newcommand\ergpcmsqps{\hbox{$\rm\thinspace erg~cm^{-2}~s^{-1}$}}
\newcommand\amsq{\hbox{$\rm\thinspace arcmin^{2}$}}
\newcommand\pdegsq{\hbox{$\rm\thinspace deg^{-2}$}}
\newcommand\msunpmpccub{\hbox{$\rm\thinspace M_{\odot}~Mpc^{-3}$}}

\begin{document}
\title*{The Obscured X-ray Background and Evolution of AGN}
\toctitle{The Obscured X-ray Background and Evolution of AGN}
%
%
\titlerunning{The Obscured X-ray Background and Evolution of AGN}
%
\author{A.C. Fabian
\and M.A. Worsley}
\authorrunning{A.C. Fabian \& M.A. Worsley}
%
%
\institute{Institute of Astronomy, Madingley Road, Cambridge CB3 0HA, UK}

\maketitle              

\begin{abstract} 

The X-ray Background has been resolved in the $0.5$--$5\kev$ band and
found to consist mostly of both unabsorbed and absorbed AGN with
column densities $<10^{23}\pcmsq$. This contrasts with the local AGN
population where the column density range extends to Compton-thick
objects and beyond ($>1.5\times 10^{24}\pcmsq$). Stacking analysis of
the integrated emission of sources detected by XMM-Newton in the
Lockman Hole, and by Chandra in the CDF-N and S reveals that the
resolved fraction of the X-ray Background drops above $6\kev$ and is
about $50\%$ above $8\kev$. The missing flux has the spectrum of
highly absorbed AGN, making it likely that the range of column
density at redshift one is similar to that locally, and that many
AGN are as yet undetected in well-studied fields.

\end{abstract}

\section{Introduction}

Deep images with Chandra and XMM-Newton have shown that the X-ray Background (XRB) is composed, at least in the $2$--$5\kev$ band, of the emission of many active galactic nuclei (AGN) \cite{giacconi02}\cite{mushotzky00}\cite{alexander03}\cite{barger02}. This was anticipated by earlier work with ROSAT \cite{hasinger98} which resolved the $0.5$--$2\kev$ XRB in a similar manner. The XRB therefore reveals the integrated X-ray emission of AGN and is an important tool for studying the accretion history of the Universe and the evolution of AGN.

The background spectrum in the $2$--$10\kev$ band is flatter than typical AGN and requires that most are absorbed \cite{setti89}\cite{madau94}\cite{comastri95}. This ties in with what we know of AGN locally where absorbed Seyfert II galaxies outnumber Seyfert Is. What remains to be determined is whether powerful objects such as quasars have a significant obscured fraction, how the obscuration evolves with time in an object and how obscuration evolves with redshift within the population.

\section{Local obscured AGN}

Absorbed AGN are very common in the local Universe \cite{matt00}. The
three nearest AGN with X-ray luminosities exceeding $10^{40}\ergps$,
NGC4945 \cite{iwasawa93}\cite{done96}, the Circinus Galaxy \cite{matt99} and Cen A \cite{rothschild99} all lie behind absorbing column densities $N_{\rm H}>10^{23}\pcmsq$. The first two are Compton thick with $N_{\rm H}>1.5\times 10^{24}\pcmsq$. The ratio of absorbed to unabsorbed nuclei is at least 3--5, with the column density distribution being fairly flat \cite{risaliti99}. The large ratio suggests that the geometry is more complex than the simple unification torus model.

At greater distances where higher luminosity objects are sampled there are some clearly highly absorbed AGN (e.g., NGC6240, \cite{vignati99}; IRAS09014, \cite{iwasawa01}; 3C294 nucleus, \cite{fabian03}). In particular, the nuclei of many radio galaxies lie behind large column densities. 

\begin{figure}[h] \begin{center}
\rotatebox{270}{\includegraphics[width=.5\textwidth]{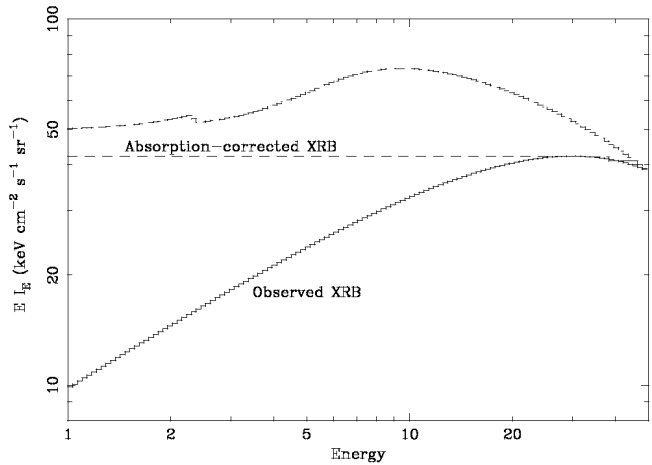}}
\end{center} \caption[]{The extragalactic XRB intensity as observed
and the minimum absorption-corrected spectrum required assuming a
typical intrinsic spectrum with $\Gamma=2$. Compton-thick obscuration
or reflection could make the corrected intensity higher.}
\label{fig1}
\end{figure}

\section {Tying the evolution of AGN and the XRB in with the local black hole mass density}

If $\eta$ is the efficiency with which mass is turned into radiation during accretion ($\eta=0.06$ for a standard thin disc around a non-spinning black hole and a typical assumed value for an accreting black hole is 0.1), then $$\varepsilon_{\rm rad} (1+z)=\eta\rho_{\rm h} c^2$$ \cite{soltan82} $\varepsilon_{\rm rad}$ is the observed energy density in that radiation now, $\rho_{\rm h}$ is the mean mass density added to the black holes and $z$ is the mean redshift of the population. Applying this formula to the {\sl intrinsic} absorption-corrected spectrum of the XRB with a 2--10~keV bolometric correction factor of 30 and a mean source redshift of 2 gave a mean black hole mass density of about $\rho_{\rm h}\sim 8\times 10^5\msunpmpccub$ \cite{fabian99}. 

This is significantly larger than $\rho_{\rm h}\sim
3\times10^5\msunpmpccub$ found by \cite{yu02} from the black hole mass
-- stellar velocity correlation applied to the local galaxy luminosity
function. Elvis et al \cite{elvis02} suggest that this requires the radiative efficiency $\eta> 0.15$ and thus that most black holes are spinning rapidly (which boosts $\eta$). 

A major new result for AGN from X-ray surveys is the peak in number
density around $z\sim 0.8$, not at $z\sim 2$ as found for quasars
\cite{hasinger02}\cite{ueda03}\cite{fiore03}. Simply put, the
population of AGN with X-ray luminosities above $3\times
10^{44}\ergps$, which are traditionally quasars, began dying at 
$z\sim 2$ whereas the population of less luminous objects, akin to
nearby Seyferts, carried on to $z\sim 0.8$ before dying. This means we
should rework the Seyfert part of the Soltan equation with a mean
$z\sim 0.8$. That changes the bolometric correction to 15
\cite{fabian03} and brings the predicted black hole mass density into
line with observations at $4-5\times 10^5\msunpmpccub$ for $\eta \sim
0.1$ \cite{fabian03}\cite{marconi04}\cite{shankar04}. 

The situation therefore looks fair for $\eta\sim 0.1$ provided that
all populations have been accounted for. A possible problem lies in
the column density distribution of local sources compared with distant
ones, such as are commonly resolved in deep X-ray surveys. The sources
found in the latter overwhelmingly have column densities
$<10^{23}\pcmsq$ \cite{alexander01} whereas locally this accounts for
only about $40\%$ \cite{risaliti99}. Does the column density
distribution evolve with redshift or have the surveys failed to find
AGN with larger column densities? To study that we have examined just
how well the spectrum of the resolved sources in deep X-ray fields
matches the spectrum of the XRB and so whether a population of sources
remains unresolved \cite{worsley04a}\cite{worsley04b}.

\section{Resolving the XRB with Deep Surveys}

The deep, pencil-beam X-ray surveys have confirmed that the XRB is
made up of point sources. The XMM-Newton observation of the Lockman
Hole (XMM-LH; \cite{hasinger01}), as well as the Chandra Deep Field North (CDF-N; \cite{alexander03}) and Chandra Deep Field South (CDF-S; \cite{giacconi02}), are able to resolve $\sim70-90\%$ of the XRB in the broad $0.5$--$2\kev$ and $2$--$10\kev$ bands. The variation in this fraction is dependent on the XRB normalisation\footnote{\cite{deluca04} summarize their own and previous work on the spectrum of the XRB which indicates that the normalization is probably a factor of 1.3 times greater than the HEAO-A2 all-sky study of \cite{marshall80}. See also \cite{revnivtsev04}.} chosen.

These high resolved fractions are often used as the basis for claims that the XRB has been completely resolved up to $10\kev$. Such optimistic statements are misleading, since it is counts in the $2$--$5\kev$ regime that dominate the $2$--$10\kev$ band. In order to investigate more carefully the behaviour of the resolved fraction with energy we carried out a source-stacking analysis using the XMM-LH data \cite{worsley04a}. We employed a straightforward photometric approach to find the total resolved flux in a number of narrow energy bands. XMM-Newton was particularly suited to the study since it has good sensitivity up to $12\kev$.

The Lockman Hole is one of the most well-studied regions of the sky and has received multi-wavelength attention. The unusually low level of neutral Galactic absorption ($\sim5\times10^{-19}\pcmsq$) makes it particularly well-suited to deep X-ray surveys. The total accumulated XMM-Newton exposure time has now reached $\sim700\ks$ from 17 different observations. We restricted our analysis to sources within a central 10-arcmin radius with the lowest background levels and maximum exposure time.

\subsection{Source Detection, Photometry and Stacking}

We cleaned the raw event files to exclude background flares and bad pixels before extracting images in five different energy bands; $0.2$--$0.5\kev$, $0.5$--$2\kev$, $2$--$4.5\kev$, $4.5$--$7.5\kev$ and $7.5$--$12\kev$. We carried out sliding-box source detection to generate an initial list of bright sources. These were then masked-out of the images and further source-fitting carried out with maximum-likelihood point spread function (PSF) fitting. The final source list was verified manually, resulting in a final list of 126 detected sources located within 10' of the observation. (Since this early work our subsequent re-analysis has made use of a more complete source list of 156 objects although the additional 30 sources make no significant difference to our results.) 

We adopted straightforward aperture photometry to determine the count-rate of each source and in each energy band. The three XMM-Newton instruments were processed separately to provide robustness and because of differences in background level and PSF. For each instrument and energy band an exposure-corrected image was made by dividing by the exposure map. In each case we also created a source-free image by masking out all the detected sources. For each source and in each energy band, the total count-rate was extracted from within a circular aperture centred on the source position. The local background level was measured in an annulus surrounding the source but here we used the source-free version of the images to avoid contamination. Given the aperture-measured and annulus-measured count-rates the true source and background count-rates can be easily determined. 

When calculating the true source and background count-rates we assumed an analytical fit to the PSF. Tests of the quality of the analytical model on a number of sources found it to be acceptable within the energy bands and off-axis angles required. The radius of the circular extraction aperture was chosen to maximise signal-to-noise in the final source count-rate. The background- and exposure-corrected count-rates for each source were converted to fluxes using conversion factors computed with \textsc{xspec} assuming a $\Gamma=1.4$ source plus $N_{\rm{H}}=5\times10^{19}\pcmsq$ Galactic absorption.

The total resolved flux in each energy band was calculated by summing the measured flux for each source. Every source is included regardless of whether or not it is detected in the band. Flux errors were added appropriately and combined with the estimated error in the counts-to-flux conversion. A correction for Galactic absorption was made; this is $\sim15\%$ in the $0.2$--$0.5\kev$ band and $\lesssim1\%$ in the harder bands.

\subsection{XMM-Newton Results}

\begin{figure}[h]
\begin{center}
\rotatebox{270}{\includegraphics[width=.5\textwidth]{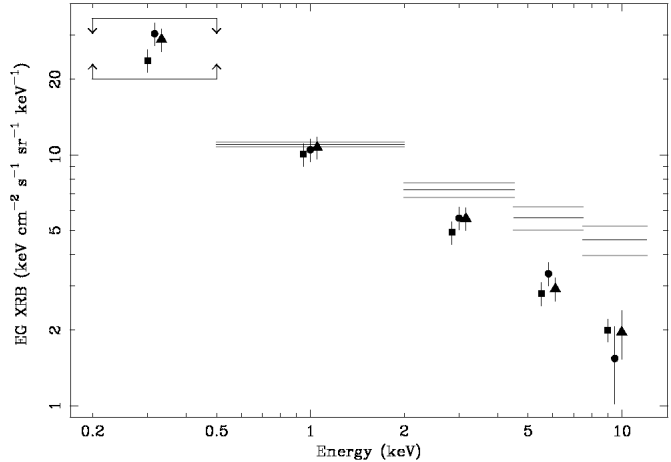}}
\end{center}
\caption[]{The extragalactic XRB intensity resolved from detected sources in the energy bands. The values from the three instruments are plotted, for clarity, as points in the centre of each band and offset horizontally with respect to each other. Squares, circles and triangles represent the PN, MOS-1 and MOS-2 instruments respectively. Errors are one sigma. The bars represent the total extragalactic XRB intensity.}
\label{fig2}
\end{figure}

Figure~\ref{fig2} shows the resolved XRB intensity for each of the XMM-Newton instruments along with the total XRB. From $0.5$--$12\kev$ we took the total XRB to be a $\Gamma=1.41$ power-law with a normalisation of $11.6\kevpcmsqpspsrpkev$ \cite{deluca04}. In the softest band, $0.2$--$0.5\kev$ the extragalactic background level is uncertain and is estimated to lie at $20$--$35\kevpcmsqpspsrpkev$ \cite{warwick98}.

From $0.2$--$2\kev$ the resolved fraction of the background is consistent with $\sim90\%$. The $2$--$4.5\kev$ band also shows a high resolved fraction of $70$--$80\%$ but the harder bands $>4.5\kev$ only resolve $\lesssim60\%$. The missing XRB component has a spectral signature consistent with what would be expected from an undetected population of highly obscured sources.

The level of the unresolved fraction depends upon the Lockman Hole being a representative sample of the sky, as well as the normalisation taken for the total XRB, which itself is still subject to errors of $10$--$15\%$. We do not correct for the bright-end source population: the pencil-beam nature of the field means that bright sources ($\gtrsim5\times10^{-14}\ergpcmsqps$) are not well-sampled whilst those brighter than $\sim10^{-13}\ergpcmsqps$ are not seen at all. The additional contribution by these bright objects would increase the resolved fraction by $\sim10$--$20\%$, but this is only comparable to the effect of field-to-field variations. Critically, since bright sources typically have very soft spectral slopes, no additional contribution can account for the missing XRB above $5\kev$, nor could any adjustments to XRB normalisation.

\subsection{The Chandra Deep Fields}

Although Chandra has low effective area at energies exceeding $7\kev$, the CDF-N and CDF-S probe much deeper than the XMM-LH. We used the CDFs to assess whether the contribution from the fainter sources detected by Chandra were able to explain the XMM-Newton results. To aid comparison, as well as to take advantage of numerous software improvements and the complete XMM-LH source catalogue, we re-analysed the XMM-Newton data using identical energy bands to the Chandra work (see \cite{worsley04b}).

The CDF-N and CDF-S cover $447.8$ and $391.3\amsq$ respectively. We used the catalogues from \cite{alexander03} which contain 503 and 326 sources respectively for the fields as well as photometry. Source-extraction apertures enclosing $90$--$100\%$ of the counts were used and the local background correction was determined from Poisson statistics. Counts-to-flux conversion was performed, where possible, using conversion factors appropriate to the hardness ratio of the source. For sources with an insufficient number of counts to permit this, a $\Gamma=1.4$ spectrum was assumed.

Prior to our stacking analysis we applied corrections for Galactic absorption as well as several systematic effects which have been recently quantified \cite{bauer04}. The total resolved flux in each of five photometric bands ($0.5$--$1\kev$, $1$--$2\kev$, $2$--$4\kev$ and $4$--$6\kev$) was calculated by summing the measured flux for each detected source. In order to calculate a resolved intensity the flux from each source must be divided by a solid angle on the sky; however, since the CDFs show a substantial increase in sensitivity towards the centre of the fields, this solid angle is a function of source brightness. The total solid angles of $447.8$ and $391.3\amsq$ are only applicable to the brightest objects, whereas fainter sources are only detectable over a fraction of these regions and the appropriate solid angle to take is smaller. We use the sensitivity functions from \cite{alexander03} when calculating source intensities.

\subsection{Bright-End Correction}

In order to remove the `noise' in the resolved fluxes that is due to poor sampling of the brighter sources (which only occur with sky densities of $\sim$ a few on the size of the deep surveys), we ignored sources with $0.5$-$8\kev$ fluxes greater than $5\times10^{-14}\ergpcmsqps$. We then used number counts from wide-area surveys to replace the missing flux due to all the sources brighter than this. We used the the $\log N$--$\log S$ functions from \cite{moretti03}, compiled in the $0.5$--$2\kev$ and $2$--$10\kev$ bands.

A number of studies have confirmed a strong dependence of source spectral shape with flux. At the bright-end sources have spectral slopes much softer than the $\Gamma=1.4$ of the XRB. We took a linear fit to the relationship as observed by \cite{streblyanska03}, although we conservatively impose a maximum of $\Gamma=2$. Given this relation we can compute the bright-source contribution for each energy band, using the appropriate spectral index at each source flux and integrating the $\log N$--$\log S$ function.

\subsection{Results}

\begin{figure}[h]
\begin{center}
\rotatebox{270}{\includegraphics[width=.5\textwidth]{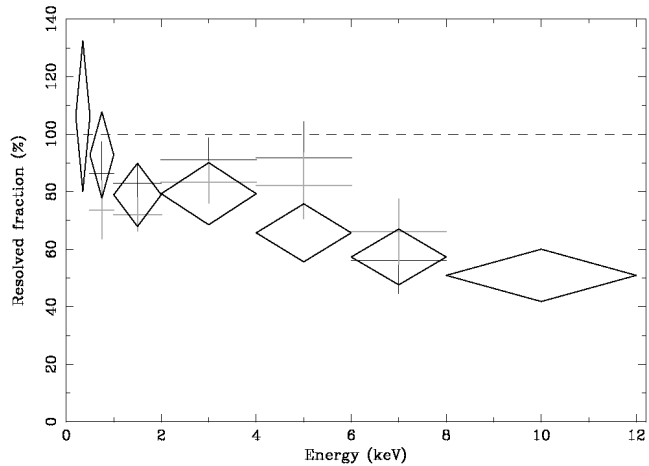}}
\end{center}
\caption[]{The resolved fraction of the XRB as seen in the CDF-N (black crosses), CDF-S (grey crosses) and XMM-LH (the combined PN/MOS-1/MOS-2 result; black diamonds), as a function of energy.}
\label{fig3}
\end{figure}

Figure~\ref{fig3} shows the bright-end corrected and  resolved XRB as seen in the CDF-N, CDF-S and XMM-LH. The extragalactic XRB level is taken to be the spectrum measured by \cite{deluca04}, but at high energies we take account of the slight turn-over, and below $1\kev$ we take account of the steeping over the spectrum implied by AGN-shadowing observations (for details refer to \cite{worsley04b}).

The resolved fraction is $\sim70$--$90\%$ up to $\sim4\kev$. XMM-Newton sees a decrease in the $4$--$6\kev$ band whereas the CDFs remain consistent with a high resolved fraction of $\sim80$--$90\%$. This difference is due to the faint, hard sources which are detected in the CDFs but not the (shallower) XMM-LH. In the $6$--$8\kev$ range all three surveys only resolve $50$--$70\%$, whilst in the $8$--$12\kev$ band XMM-Newton only resolves $40$-$60\%$ of the background. The total resolved flux in the broad $0.5$--$8$ range remains high ($80$--$85\%$), and illustrates the need for narrow-band stacking to reveal the decrease $>6\kev$.

\subsection{The Missing Source Population}

\begin{figure}[h]
\begin{center}
\rotatebox{270}{\includegraphics[width=.5\textwidth]{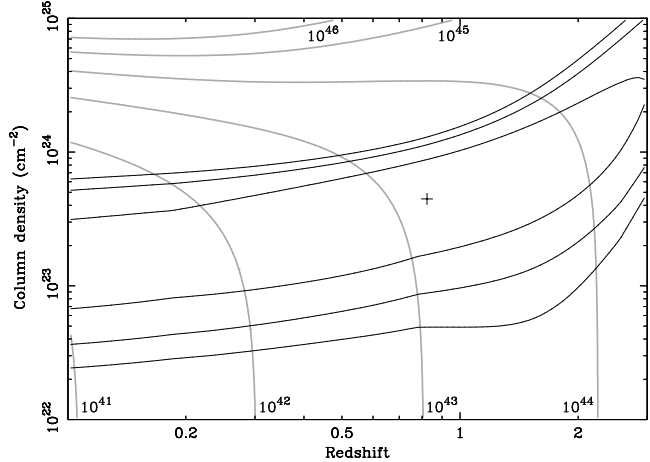}}
\end{center}
\caption[]{Confidence contours (horizontal) of fits between the
missing XRB spectrum and an obscured power-law model for the
XMM-Newton data. 68, 90 and 95\% contours are shown. The best-fitting
model has $z\sim0.8$ and $N_{\rm{H}}\sim4.5\times10^{23}\pcmsq$. The
other contours indicate the maximum, unobscured, rest-frame $2$--$10\kev$ luminosity of a single source in the population.}
\label{fig4}
\end{figure}

We have modelled the spectral shape of the missing XRB fraction as a single population of sources. We assumed an underlying $\Gamma=2$ power-law, inclusive of an $R=1$ reflection component, plus intrinsic photoelectric absorption. A grid of spectra was constructed over the range $z=0.1$--$3$ and $N_{\rm{H}}=10^{22}$--$10^{25}\pcmsq$ and for each we calculated the goodness-of-fit between the model spectrum and the unresolved XRB spectrum. The confidence contours in source spectral shape are shown in Figure~\ref{fig4}.

\begin{figure}[h]
\begin{center}
\rotatebox{270}{\includegraphics[width=.5\textwidth]{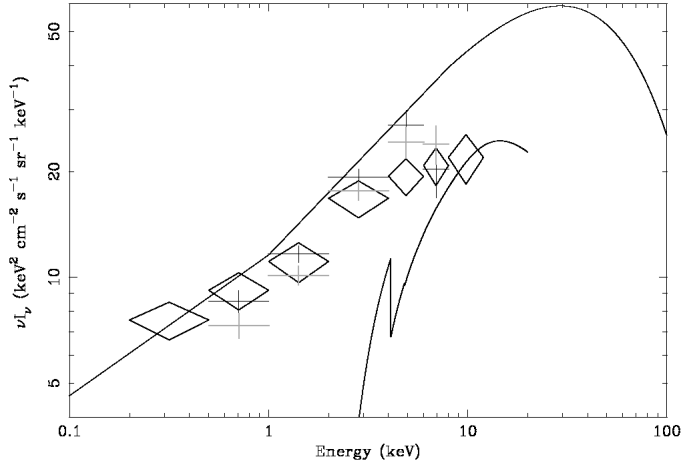}}
\end{center}
\caption[]{The total extragalactic XRB (upper solid curve) along with that resolved in the XMM-LH (black diamonds), CDF-N (black crosses) and CDF-S (grey crosses). The lower solid curve shows the spectrum of the best-fitting model for the missing source population at $z\sim0.8$ and with $N_{\rm{H}}\sim4.5\times10^{23}\pcmsq$.}
\label{fig5}
\end{figure}

The XMM-LH and CDFs show similar contours for the characteristics of
the missing population, although it is the XMM-Newton results, which
extend up to $12\kev$, which provide the best constraints. The
best-fitting models have $z=0.5$--$1.5$ and
$N_{\rm{H}}=10^{23}$--$10^{24}\pcmsq$, pointing to a highly obscured
population of AGN. Since the missing sources are undetected, the
survey sensitivity allows an upper limit to be placed on the
luminosity of each object. For the best-fitting source models this
upper limit is $\lesssim5\times10^{43}\ergps$. The sources must occur
with areal densities of $\gtrsim2800\pdegsq$ -- a considerable number.
This would correspond to at least $\sim350$ undetected AGN in the
CDF-N, and suggests that there must be $\gtrsim3$ times more undetected,
obscured sources than detected, unobscured ones. Figure~\ref{fig5}
shows the resolved and total XRB with the typical shape of the sources
required to account for the missing intensity.

The absorption-corrected flux of the missing population can be used to estimate the black hole mass density of the obscured population. Assuming an accretion efficiency of 0.1, and a bolometric correction of 0.1, the missing population would correspond to a black hole mass density of $\sim0.5$--$1\times10^{5}\msunpmpccub$; $\sim10$--$20\%$ of the total local value.

\section{Summary}

Studies of the XRB continue to reveal the evolution and properties of AGN. Most of the XRB is resolved up to an energy of 5~keV and is due to AGN -- Seyferts and quasars -- with absorbing column densities $<2\times 10^{23}\pcmsq$. The identification of large samples of such objects shows that the lower luminosity ones (Seyfert-like) evolve rapidly to $z\sim 0.8$ while quasars evolve more slowly to $z\sim 2$.  Assuming that the Seyfert-like ones have lower black hole masses, those black holes continued to grow at later stages of the Universe than the more massive ones associated with quasars. 

The resolved fraction of the XRB drops to $60$--$70\%$ above 6~keV and
$50\%$ above 8~keV. The missing fraction between 6--8~keV is not due
to the known faint sources in the 2--5~keV band since it would have
been detected by the stacking analysis which corresponds to an
exposure time of 0.5~Gs or more.  The missing fraction is most simply
explained as due to the high column density ($N_{\rm H}>2\times
10^{23}\pcmsq$) extension to the observed Seyfert-like objects
evolving rapidly to $z\sim 1$. Such a population was implicit in XRB
synthesis models such as that of \cite{gandhi03} or \cite{gilli01}.

Presumably there is also a Compton-thick population as well, which
would make less of an impact on the XRB since Compton down-scattering
then depletes the emergent X-ray luminosity. The total black
hole density due to Compton-thin sources is then about $\rho_{\rm
h}\approx 5\times 10^5 \eta^{-1}_{-1}\msunpmpccub$, where
$\eta=0.1\eta_{-1}$. This is consistent with recent estimates of the
local value of $\rho_{\rm h}$ \cite{marconi04}\cite{shankar04} for
$\eta=0.1$. 

The issue of highly obscured and Compton-thick quasars remains
unresolved \cite{comastri04}. Powerful radio galaxies represent one,
relatively small, population. Some powerful, highly obscured,
radio-quiet quasars have been found
\cite{crawford02}\cite{norman02}\cite{stern02}\cite{wilman03}, with
\cite{gandhi04} estimating a surface density of at least
10~deg$^{-2}$, The uncertainties are currently too large to assess
their general importance. Future work combining X-ray and Spitzer data
will be instructive.

%


\begin{thebibliography}{8.}
\addcontentsline{toc}{section}{References}

\bibitem{giacconi02} R.~Giacconi, A.~Zirm, J.~Wang, P.~Rosati, M.~Nonino, P.~Tozzi, R.~Gilli, V.~Mainieri, G.~Hasinger, L.~Kewley, J.~Bergeron, S.~Borgani, R.~Gilmozzi, N.~Grogin, A.~Koekemoer, E.~Schreier, W.~Zheng, C.~Norman: ApJS \textbf{139}, 369 (2002)

\bibitem{mushotzky00} R.F.~Mushotzky, L.L.~Cowie, A.J.~Barger, K.A.~Arnaud: Nat \textbf{404}, 459 (2000)

\bibitem{alexander03} D.M.~Alexander, F.E.~Bauer, W.N.~Brandt, D.P.~Schneider, A.E.~Hornschemeier, C.~Vignali, A.J.~Barger, P.S.~Broos, L.L.~Cowie, G.P.~Garmire, L.K.~Townsley, M.W.~Bautz, G.~Chartas, W.L.W.~Sargent: AJ \textbf{126}, 539 (2003)

\bibitem{barger02} A.J.~Barger, L.L.~Cowie, W.N.~Brandt, P.~Capak, G.P.~Garmire, A.E.~Hornschemeier, A.T.~Steffen, E.H.~Wehner: AJ \textbf{124}, 1839 (2002)

\bibitem{hasinger98} G.~Hasinger, R.~Burg, R.~Giacconi, M.~Schmidt, J.~Trumper, G.~Zamorani: A\&A \textbf{392}, 482 (1998)

\bibitem{setti89} G.~Setti, L.~Woltjer: A\&A \textbf{224}, L21 (1989)

\bibitem{madau94} P.~Madau, G.~Ghisellini, A.C.~Fabian: MNRAS \textbf{270}, L17 (1994)

\bibitem{comastri95} A.~Comastri, G.~Setti, G.~Zamorani, G.~Hasinger A\&A \textbf{296}, 1 (1995)

\bibitem{matt00} G.~Matt, A.C.~Fabian, M.~Guainazzi, K.~Iwasawa, L.~Bassani, G.~Malaguti: MNRAS \textbf{318}, 173 (2000)

\bibitem{iwasawa93} K.~Iwasawa, K.~Koyama, H.~Awaki, H.~Kunieda, K.~Makishima, T.~Tsuru, T.~Ohashi, N.~Nakai: ApJ \textbf{409}, 155 (1993)

\bibitem{done96} C.~Done, G.M.~Madejski, D.A.~Smith: ApJ \textbf{463}, L63 (1996)

\bibitem{matt99} G.~Matt, M.~Guainazzi, R.~Maiolino, S.~Molendi, G.C.~Perola, L.A.~Antonelli, L.~Bassani, W.N.~Brandt, A.C.~Fabian, F.~Fiore, K.~Iwasawa, G.~Malaguti, A.~Marconi, J.~Poutanen: A\&A \textbf{341}, L39 (1999)

\bibitem{rothschild99} R.E.~Rothschild, D.L.~Band, P.R.~Blanco, D.E.~Gruber, W.A.~Heindl, D.R.~MacDonald, D.C.~Marsden, K.~Jahoda, D.~Pierce, G.~Madejski, M.~Elvis, D.A.~Schwartz, R.~Remillard, A.A.~Zdziarski, C.~Done, R.~Svensson: ApJ \textbf{510}, 651 (1999)

\bibitem{risaliti99} G.~Risaliti, R.~Maiolino, M.~Salvati: ApJ \textbf{522}, 157 (1999)

\bibitem{vignati99} P.~Vignati, S.~Molendi, G.~Matt, M.~Guainazzi, L.A.~Antonelli, L.~Bassani, W.N.~Brandt, A.C.~Fabian, K.~Iwasawa, R.~Maiolino, G.~Malaguti, A.~Marconi, G.C.~Perola: A\&A \textbf{349}, L57 (1999) 

\bibitem{iwasawa01} K.~Iwasawa, A.C.~Fabian, S.~Ettori: MNRAS \textbf{321}, L15 (2001)

\bibitem{fabian03} A.C.~Fabian: `Carnegie Observatories Astrophysics Series, Vol. 1: Coevolution of Black Holes and Galaxies' ed. L.C.~Ho (Cambridge: Cambridge Univ. Press) (astro-ph/0304122) (2003)

\bibitem{soltan82} A.~Soltan: MNRAS \textbf{200}, 115 (1982)

\bibitem{fabian99} A.C.~Fabian, K.~Iwasawa: MNRAS \textbf{303}, L34 (1999)

\bibitem{yu02} Q.~Yu, S.~Tremaine: MNRAS \textbf{335}, 965 (2002)

\bibitem{elvis02} M.~Elvis, G.~Risalti, G.~Zamorani: ApJ \textbf{565}, L75 (2002)

\bibitem{hasinger02} G.~Hasinger: Proc. Symposium `New Visions of the X-ray Universe in the XMM-Newton and Chandra Era' 26-30. November, ESTEC, The Netherlands (2001)

\bibitem{ueda03} Y.~Ueda, M.~Akiyama, K.~Ohta, T.~Miyaji: ApJ \textbf{598}, 886 (2003)

\bibitem{fiore03} F.~Fiore, M.~Brusa, F.~Cocchia, A.~Baldi, N.~Carangelo, P.~Ciliegi, A.~Comastri, F.~La~Franca, R.~Maiolino, G.~Matt, S.~Molendi, M.~Mignoli, G.C.~Perola, P.~Severgnini, C.~Vignali: A\&A \textbf{409} 79 (2003)

\bibitem{marconi04} A.~Marconi, G.~Risaliti, R.~Gilli, L.K.~Hunt, R.~Maiolino, M.~Salvati: MNRAS \textbf{351} 169 (2004)

\bibitem{shankar04} F.~Shankar, P.~Salucci, G.L.~Granato, G.~De~Zotti, L.~Danese: MNRAS, in press (astro-ph/0405585)

\bibitem{alexander01} D.M.~Alexander, W.N.~Brandt, A.E.~Hornschemeier, G.P.~Garmire, D.P.~Schneider, F.E.~Bauer, R.E.~Griffiths: AJ \textbf{122}, 2156 (2001) 

\bibitem{worsley04a} M.A.~Worsley, A.C.~Fabian, X.~Barcons, S.~Mateos, G.~Hasinger, H.~Brunner: MNRAS \textbf{352}, L28 (2004)

\bibitem{worsley04b} M.A.~Worsley, A.C.~Fabian, F.E.~Bauer,
D.M.~Alexander, G.~Hasinger, S.~Mateos, H.~Brunner, W.N.~Brandt,
D.P.~Schneider: MNRAS, submitted (2004)

\bibitem{hasinger01} G.~Hasinger, B.~Altieri, M.~Arnaud, X.~Barcons, J.~Bergeron, H.~Brunner, M.~Dadina, K.~Dennerl, P.~Ferrando, A.~Finoguenov, R.E.~Griffiths, Y.~Hashimoto, F.A.~Jansen, D.H.~Lumb, K.O.~Mason, S.~Mateos, R.G.~McMahon, T.~Miyaji, F.~Paerels, M.J.~Page, A.F.~Ptak, T.P.~Sasseen, N.~Schartel, G.P.~Szokoly, J.~Tr\" umper, M.~Turner, R.S.~Warwick, M.G.~Watson: A\&A \textbf{365}, L45 (2001)

\bibitem{deluca04} A.~De~Luca, S.~Molendi: A\&A \textbf{419}, 837 (2004)

\bibitem{marshall80} F.E.~Marshall, E.A.~Boldt, S.S.~Holt, R.B.~Miller, R.F.~Mushotzky, L.A.~Rose, R.E.~Rothschild, P.J.~Serlemitsos: ApJ \textbf{235}, 4 (1980)

\bibitem{revnivtsev04} M.~Revnivtsev, M.~Gilfanov, R.~Sunyaev, K.~Jahoda, C.~Markwardt: A\&A \textbf{411}, 329 (2003)

\bibitem{warwick98} R.S.~Warwick, T.P.~Roberts: AN \textbf{319}, 59 (1998)

\bibitem{bauer04} F.E.~Bauer, D.M.~Alexander, W.N.~Brandt, D.P.~Schneider, E.~Treister, A.E.~Hornschemeier, G.P.~Garmire: AJ, in press (astro-ph/0408001) (2004)

\bibitem{moretti03} A.~Moretti, S.~Campana, D.~Lazzati, D.~Tagliaferri: ApJ \textbf{588}, 696 (2003)

\bibitem{streblyanska03} A. Streblyanska, J. Bergeron, H. Brunner, A.
Finoguenov, G. Hasinger, V. Mainieri: Nucl. Phys. B., Vol. 132, {\it
Proceedings of the 2nd BeppoSAX Conference: The Restless High-Energy
Universe}, Eds. E.P.J. van den Heuvel, R.A.M.J. Wijers, J.J.M. in 't
Zand, p. 232 (2004)



\bibitem{gandhi03} P.~Gandhi, A.C.~Fabian: MNRAS \textbf{339}, 1095 (2003)

\bibitem{gilli01} R.~Gilli, M.~Salvati, G.~Hasinger: A\&A \textbf{366}, 407 (2001)

\bibitem{comastri04} A.~Comastri, F.~Fiore, C.~Vignali, M.~Brusa,
F.~Civano: these proceedings. (astro-ph/0410272) (2004)

\bibitem{crawford02} C.S.~Crawford, P.~Gandhi, A.C.~Fabian, R.J.~Wilman, R.M.~Johnstone, A.J.~Barger, L.L.~Cowie: MNRAS \textbf{333}, 809 (2002)

\bibitem{norman02} C.~Norman, G.~Hasinger, R.~Giacconi, R.~Gilli, L.~Kewley, M.~Nonino, P.~Rosati, G.~Szokoly, P.~Tozzi, J.~Wang, W.~Zheng, A.~Zirm, J.~Bergeron, R.~Gilmozzi, N.~Grogin, A.~Koekemoer, E.~Schreier: ApJ \textbf{571}, 218 (2002)

\bibitem{stern02} D.~Stern, E.C.~Moran, A.L.~Coil, A.~Connolly, M.~Davis, S.~Dawson, A.~Dey, P.~Eisenhardt, R.~Elston, J.R.~Graham, F.~Harrison, D.J.~Helfand, B.~Holden, P.~Mao, P.~Rosati, H.~Spinrad, S.A.~Stanford, P.~Tozzi, K.L.~Wu: ApJ \textbf{568}, 71 (2002)

\bibitem{wilman03} R.J.~Wilman, A.C.~Fabian, C.S.~Crawford, R.M.~Cutri: MNRAS \textbf{338}. L19 (2003)

\bibitem{gandhi04} P.~Gandhi, C.S.~Crawford, A.C.~Fabian, R.M.~Johnstone: MNRAS \textbf{348}, 529 (2004)

\end{thebibliography}
\end{document}